\documentclass[superscriptaddress,showpacs,twocolumn,aps,pre,10pt]{revtex4-1}

\usepackage{graphicx,dcolumn,bm,amsmath,color,amssymb,relsize}
\usepackage{units}    
\usepackage[utf8]{inputenc}
\usepackage{ dsfont }
\usepackage{todonotes}

\newcommand{\fig}[1]{Fig.~\ref{#1}}
\newcommand{\avg}[1]{ {\langle #1 \rangle} }

\newcommand{\shortint}{\hspace{-.1em}\int\hspace{-.25em}}

\newcommand{\br}{ {\bf r} }
\newcommand{\bR}{ {\bf R} }

\newcommand{\bF}{ {\bf F} }
\newcommand{\bff}{ \boldsymbol{f} } 
 
\newcommand{\kbt}{k_{\rm B}T}

\newcommand{\D}{ {\rm d} }

\newcommand{\ga}{ {\alpha }}
\newcommand{\gb}{ {\beta }}
\newcommand{\gc}{ {\gamma }}
\newcommand{\gas}{ {\alpha^\prime }}
\newcommand{\gbs}{ {\beta^\prime }}

\newcommand{\ddt}[1]{\frac{\partial #1}{\partial t}}
\newcommand{\PP}{\ensuremath{\Psi(\bR, t)}}

\begin{document}

\title{Structural correlations in diffusiophoretic colloidal mixtures with nonreciprocal interactions }

\author{J\"org Bartnick}
\email{bartnick@thphy.uni-duesseldorf.de}
\affiliation{Institut f\"ur Theoretische Physik II: Weiche Materie, Heinrich-Heine-Universit\"at D\"{u}sseldorf, 
D-40225 D\"{u}sseldorf, Germany}

\author{Marco Heinen}
\affiliation{Division of Chemistry and Chemical Engineering, California Institute of Technology, Pasadena, California 91125, USA}

\author{Alexei V. Ivlev}
\affiliation{Max-Planck-Institut f\"ur Extraterrestrische Physik, 85741 Garching, Germany}

\author{Hartmut L\"owen}
\affiliation{Institut f\"ur Theoretische Physik II: Weiche Materie, Heinrich-Heine-Universit\"at D\"{u}sseldorf, 
D-40225 D\"{u}sseldorf, Germany}

\date{\today}

\begin{abstract}
Nonreciprocal effective interaction forces can occur between mesoscopic particles in colloidal suspensions that are driven out of equilibrium.
These forces violate Newton's third law \mbox{\it actio=reactio\/} on coarse-grained length and time scales.
Here we explore the statistical mechanics of Brownian particles with nonreciprocal effective interactions.
Our model system is a binary fluid mixture of spherically symmetric, diffusiophoretic mesoscopic particles, and we focus on the time-averaged particle pair- and
triplet-correlation functions. Based on the many-body Smoluchowski equation we develop a microscopic statistical theory
for the particle correlations and test it by computer simulations. For model systems in two and three spatial dimensions,
we show that nonreciprocity induces distinct nonequilibrium pair correlations.  
Our predictions can be tested in experiments with chemotactic colloidal suspensions.
\end{abstract}


\maketitle

\section{Introduction}
Mesoscopic Brownian particles in colloidal suspensions, with typical particle diameters between a few nanometers and a few micrometers,
exert forces on each other that depend on the microscopic position and velocity variables of many molecules in the suspending solvent.
On coarse-grained length and time scales where the solvent microstructure and dynamics are not resolved, the solvent molecule's degrees
of freedom can be `integrated out' and one is left with colloidal particles that interact via \emph{effective} forces. These effective
forces depend on the thermodynamic state of the solvent. 
The tunability of the effective interactions between colloidal particles makes colloidal suspensions ideal model systems for
studying classical many-body behavior such as crystallization \cite{Poon2002,Kahl2009,Palberg2014,Pusey1991},
melting \cite{Murray1996,Zhu1997,Lowen1994}, phase separation \cite{Lekkerkerker1992,Adams1998,Stradner2004,Lowen1997}
as well as glass and gel formation \cite{Hunter2012,Royall2007,Lu2008,Lowen2012}. In thermodynamic equilibrium, the
effective interactions fulfill Newton's third law {\it actio=reactio\/}. That is: the effective force generated by a particle
and acting on a second particle is equal in magnitude and opposite in direction, when compared to the force generated by
the second particle, acting on the first particle \cite{Israelachvili1995,Dijkstra2000,Bolhuis2001}.
  
However, the {\it actio=reactio\/} principle can be broken in a nonequilibrium situation. 
Nonreciprocity occurs in a multitude of systems. Naming a few examples only, nonreciprocity
can arise from nonequilibrium fluctuations \cite{Hayashi2006,Buenzli2008} and also
in case of diffusiophoretic forces~\cite{soto2014,soto2015}, optical forces \cite{Dholakia2010,Shanblatt2011}, out-of-equilibrium depletion interactions \cite{Khair2007,Dzubiella2003,Sriram2012},
hydrodynamic interactions \cite{Alabrudzinski2009}, and `social forces' in pedestrian dynamics modeling \cite{Helbing1995,Helbing2000}.
Nonreciprocal effective interactions are typically superimposed by
the classical reciprocal interactions, stemming from electric charges or dipole moments on the particles, van der Waals interactions,
excluded volume, or other types of direct interactions.

Despite their importance, the many-body statistics of particles with
nonreciprocal interactions have not been studied so far in the context of colloidal suspensions. 
This stands in stark contrast to the topic of complex (dusty) plasmas \cite{Lowen2012}, where nonreciprocal interactions are a familiar feature
of anisotropic trailing space-charges in the downstream direction behind charged mesoscopic particles in a flowing plasma. The phenomenon is known as the plasma wake. 
Consequences of nonreciprocity have been explored in various studies concerning complex plasmas~\cite{Ivlev2000,Couedel2010,Schweigert2000,Qiao2013,Qiao2014,Melzer1999,Steinberg2001}.
The most prominent difference between colloidal suspensions and complex plasmas is that the dynamics of colloidal particles in high-density viscous solvent
is completely overdamped while the dust-grain dynamics in complex plasmas typically contain a large inertial contribution.

The binary colloidal model system that we study in this paper is governed by pairwise additive nonreciprocal forces and erratic Brownian forces.
Like in Ref.~\cite{Ivlev2015}, we characterize the strength of nonreciprocity by a scalar parameter $\Delta$ which is the ratio of the nonreciprocal to reciprocal forces.
We focus on the time-averaged pair- and triplet-correlation functions for particle positions, developing a microscopic statistical theory based on the many-body-Smoluchowski equation
and the Kirkwood superposition approximation as a closure [see Eq.~\eqref{kirkwood}]. The theory is successfully tested against our Brownian dynamics computer simulations.
As a result, we find that nonreciprocity induces distinct nonequilibrium pair correlations, and we also analyze the triplet correlations and the impact of the
Kirkwood superposition approximation.

\section{The Model} 
\label{sec:model} 

Our model system is a generalized variant of a diffusiophoretic particle suspension that has been studied by Soto and Golestanian \cite{soto2014}.
Consider an equimolar Brownian suspension containing two different types, $A$ and $B$, of spherically symmetric, catalytic mesoscopic particles.
We denote the time-depended position of particle $i$ of type $\alpha$ by the row vector $\boldsymbol{r}_i^\alpha(t)$.
The suspension contains $2N$ particles, and we define the super vector
\begin{equation}
  \bR(t) = \left( \br^A_1(t), \ldots, \br^A_{N}(t), \br^B_1(t), \ldots, \br^B_{N}(t) \right)\nonumber
\end{equation}
as short-hand notation for the positions of all particles.
Throughout this paper, upper  indices containing Greek or capital Roman letters are species indices that should not be confused with exponents. 

Let each particle of type $A$ act as a source of strength $s_A$, for a chemical substance $\mathcal{A}$ that consists of small molecules.
Likewise, let each particle of type $B$ be a source of strength $s_B$ for a low molecular weight chemical substance $\mathcal{B}$.
The molecules of substances $\mathcal{A}$ and $\mathcal{B}$ undergo diffusive motion in the solvent phase, characterized by
the Stokes-Einstein-Sutherland translational diffusion coefficients $D_{\mathcal{A}}$ and $D_{\mathcal{B}}$, respectively.
Evaporation or chemical decomposition into inert products causes molecules of types $\mathcal{A}$ and $\mathcal{B}$
to disappear at constant rates $\nu_{\mathcal{A}}$ and $\nu_{\mathcal{B}}$, respectively~\cite{Grima2005,Grima2006}.
The explicitly position- and time-dependent concentration fields of the two chemical substances, $c_{\mathcal{A}}(\br,t)$ and $c_{\mathcal{B}}(\br,t)$,
depend in general also on $\bR(\tau)$ at all times $\tau < t$. However, we assume that the diffusion coefficients $D_{\mathcal{A}}$ and $D_{\mathcal{B}}$
are large enough to allow for a separation of time scales: At a coarse-grained time scale, each individual particle traverses a
distance that is much smaller than the average distance to the nearest neighboring particle and the particle configuration $\bR$ is therefore
practically unchanged. At the same time scale, the fast diffusion of
$\mathcal{A}$- and $\mathcal{B}$-type molecules has already led to steady-state concentration fields $c_{\mathcal{A}}(\br,t)$ and $c_{\mathcal{B}}(\br,t)$
that depend only on the instantaneous particle positions $\bR(t)$, but not on the history $\bR(\tau)$~\cite{DeBuyl2013}.  
Restricting our study to time scales that are longer than the mentioned coarse-grained time scale, and neglecting 
all direct correlations between the two chemical substances' molecules, the concentration fields are governed by the 
instantaneous diffusion equations%
\begin{equation}
\nu_{\mathcal{A}}c_{\mathcal{A}}(\br,t) - D_{\mathcal{A}}~\nabla^2 c_{\mathcal{A}}(\br,t) =
s_A ~\sum_{i=1}^{N}\delta(\br - \boldsymbol{r}_i^A(t)) \label{eq:diffA}
\end{equation}
and
\begin{equation}
\nu_{\mathcal{B}}c_{\mathcal{B}}(\br,t) - D_{\mathcal{B}}~\nabla^2 c_{\mathcal{B}}(\br,t) =
s_B ~\sum_{i=1}^{N}\delta(\br - \boldsymbol{r}_i^B(t)), \label{eq:diffB}
\end{equation}
where $\nabla^2$ is the Laplace operator with respect to the field point $\br$ and $\delta(\br)$ is the Dirac delta function.
For the sake of simplicity we approximate the particles as point-like objects, as reflected by the point sources on the right-hand sides of
Eqs.~\eqref{eq:diffA} and \eqref{eq:diffB}. This point-particle approximation is justified if the typical distances between particles
are much larger than the particle diameters, which is the case for the systems that we have studied (see Fig.~\ref{fig1} and the relating text in Sec.~\ref{sec:paircorrelation}).

Solving the linear screened Poisson equations \eqref{eq:diffA} and \eqref{eq:diffB}
by standard Green's function methods gives the result
\begin{equation}\label{eq:concfieldA}
c_{\mathcal{A}}(\br,t) = \frac{s_A}{D_{\mathcal{A}}} \sum\limits_{j=1}^{N} G\left(\sqrt{\frac{D_{\mathcal{A}}}{\nu_{\mathcal{A}}} }, |\br - \br_j^A(t)|\right)  
\end{equation}
and an analogous expression for $c_{\mathcal{B}}(\br,t)$ which is obtained after the interchange of indices $\mathcal{A} \to \mathcal{B}$ and $A \to B$.
In Eq.~\eqref{eq:concfieldA}, $G(\lambda, r) = \exp(-r/\lambda) / (4\pi r)$ is the isotropic Green's function in terms of the norm $r = |\br|$ of
vector $\br$, satisfying the equation $(\nabla^2 - \lambda^{-2}) G(\lambda,r) = - \delta(\br)$ with an exponential screening length $\lambda$.
Nonzero values of $\nu_{\mathcal{A}}$ and $\nu_{\mathcal{B}}$ correspond to finite values for $\lambda$, which
sets our model apart from unscreened chemotatic models with zero evaporation rate and $\lambda \to \infty$~\cite{Tsori2004}.

We continue by picking an arbitrary tagged particle $i$ of species $A$, and splitting the sum in Eq.~\eqref{eq:concfieldA}
into a self-part $(i=j)$ and a complementary distinct part $(i \neq j)$. The self-part gives the concentration field
\begin{equation}\label{eq:selfconcfieldA}
c_{\mathcal{A},i}^s(\br,t) = \frac{s_A}{D_{\mathcal{A}}} G\left(\sqrt{\frac{D_{\mathcal{A}}}{\nu_{\mathcal{A}}} }, |\br - \br_i^A(t)|\right)  
\end{equation}
of chemical $\mathcal{A}$, which is created by the tagged particle around itself, and which is isotropic around $\br = \br_i^A$.
The anisotropic distinct part
\begin{equation}\label{eq:distinctconcfieldA}
c_{\mathcal{A},i}^d(\br,t) =
\frac{s_A}{D_{\mathcal{A}}} \sum\limits_{\substack{j=1 \\ j \neq i}}^{N} G\left(\sqrt{\frac{D_{\mathcal{A}}}{\nu_{\mathcal{A}}} }, |\br -  \br_j^A(t)|\right)  
\end{equation}
is created by the remaining particles of species $A$ and, obviously, $c_{\mathcal{A}}(\br,t) = c_{\mathcal{A},i}^s(\br,t) + c_{\mathcal{A},i}^d(\br,t)$.
Once again, Eqs.~\eqref{eq:selfconcfieldA} and \eqref{eq:distinctconcfieldA} can be repeated analogously for the chemical species $\mathcal{B}$ and a tagged
particle of species $B$, by interchange of indices $\mathcal{A} \to \mathcal{B}$ and $A \to B$.

Diffusiophoretic particles tend to drift in the direction parallel or opposite to a chemical substance's concentration gradient \cite{Anderson1989,Sabass2010,Brady2011,Sengupta2011}.
Assuming concentration- and configuration-independent mobility coefficients $\mu_{A\mathcal{A}}$, $\mu_{A\mathcal{B}}$, $\mu_{B\mathcal{A}}$ and $\mu_{B\mathcal{B}}$, with dimension $\text{Force} \times \text{Length}^4$, 
we define the total diffusiophoretic forces
\begin{equation}\label{eq:diffu_force_on_A}
\bF_i^A (\bR) =   - \mu_{A\mathcal{A}} {\left.\nabla c_{\mathcal{A},i}^d(\br)\right|}_{\br=\br_i^A}
- \mu_{A\mathcal{B}} {\left.\nabla c_{\mathcal{B}}(\br)\right|}_{\br=\br_i^A}
\end{equation}
and
\begin{equation}\label{eq:diffu_force_on_B}
\bF_j^B (\bR) =  - \mu_{B\mathcal{A}} {\left.\nabla c_{\mathcal{A}}(\br)\right|}_{\br=\br_j^B}  
- \mu_{B\mathcal{B}} {\left.\nabla c_{\mathcal{B},j}^d(\br)\right|}_{\br=\br_j^B}
\end{equation}
acting on particle $i$ of species $A$, and on particle $j$ of species $B$, respectively. 
In Eqs.~\eqref{eq:diffu_force_on_A} and \eqref{eq:diffu_force_on_B} we drop the instantaneous time-dependence for clarity.
Note that the diffusiophoretic force on a particle is not affected by the isotropic self-part of the concentration field around the respective particle, but only
by the distinct part of the concentration fields, created by all other particles.
This is analogous to the forces among a set of point-like electric charges, \textit{e.g.} electrons: The Lorentz force on a single electron depends on the
positions and velocities of all other electric charges, but it is not affected by the field that the tagged particle creates itself.

Our model shares many properties with a system of electric point charges that interact via pairwise additive screened Coulomb forces,
like charged particles moving in an electrolyte, with electric fields calculated in the Debye-Hückel approximation.
Combining Eqs.~\eqref{eq:concfieldA} and \eqref{eq:distinctconcfieldA}-\eqref{eq:diffu_force_on_B},
we can interpret the individual summands that contribute to $\bF_i^A (\bR)$ and $\bF_j^B (\bR)$ 
as pairwise additive forces $\bF^{\alpha\beta}(\br_j^\beta - \br_i^\alpha)$, exerted by particle $i$ of species $\alpha$ on particle $j$ of species $\beta$.
However, a peculiarity of the binary diffusiophoretic particle mixture that sets it qualitatively apart from the ensemble of electric point charges
is the action-reaction symmetry breaking: An inequality 
\begin{equation}
 \bF^{AB}(\br_j^B - \br_i^A) \neq - \bF^{BA}( \br_i^A - \br_j^B) \label{eq:nonrecip_pair_forces}
\end{equation}
occurs in the general case and, as in Ref.~\cite{Ivlev2015}, we introduce a scalar nonreciprocity parameter $\Delta(r)$ by the defining equation
\begin{equation}\label{eq:Delta_definition}
\Delta(r) \left[\bF^{AB}(r) + \bF^{BA}(r)\right] = \bF^{BA}(r) - \bF^{AB}(r).
\end{equation} 
In the reciprocal case, where $\bF^{AB}(r) = \bF^{BA}(r)$, this parameter vanishes and we have $\Delta = 0$.

In the following we neglect hydrodynamic interactions, which can be justified if the suspension is highly dilute but 
still strongly interacting. For our analysis to be valid, the hydrodynamic diameters of the particles have to be
much smaller than the shortest typical particle distances in the suspension. Particles with sufficiently strong, repulsive
Yukawa-like interactions virtually never come into close contact, and the characteristic length scale that dominates
the correlation functions of such particles in $d$-dimensional space is $\rho^{-1/d}$, where $\rho$ is the particle
number density \cite{Westermeier2012,Heinen2014}. Suspensions of such particles can exhibit strong structural correlations,
even if they are highly dilute from a hydrodynamic point of view.

The Brownian particle dynamics, on time scales that exceed the momentum
relaxation time, are described by the overdamped Langevin equation \cite{Doi1986}
\begin{equation}
\xi^\ga ~ \dot{\br}_i^\ga =\bF ^\ga _i (\bR,t) + \bff_i^\ga(t)\label{langevin} 
\end{equation}
with a friction coefficient $\xi^\ga$ and a random force $\bff_i^\ga(t)$ with zero mean, $\avg{\bff_i^\ga(t)} = 0$, and variance
$\avg{\bff_i^\ga(t)\bff_j^\gb(\tau)} = 2 \kbt \xi^\ga  \delta_{ij}  \delta_{\ga\gb} \delta(t- \tau) \mathds{1}$.
Here, $\delta_{ij}$ is the Kronecker symbol, $\mathds{1}$ is the unit matrix,
the brackets $\avg{\vphantom{\bff_i^\ga}\dots}$ represent an average with respect to the time $t$,
and $k_B$ and $T$ denote Boltzmann's constant and absolute temperature, respectively.

Note here that our model system makes minimal assumptions about the nature of the diffusiophoretic particles only:
Particles are fully characterized by their monopolar source terms $s_A$ and $s_B$, their diffusiophoretic mobilities
$\mu_{A\mathcal{A}}$, $\mu_{A\mathcal{B}}$, $\mu_{B\mathcal{A}}$ and $\mu_{B\mathcal{B}}$, and their friction coefficients $\xi^A$ and $\xi^B$.
The particle environment is fully described by the chemical substance diffusion coefficients $D_{\mathcal{A}}$ and $D_{\mathcal{B}}$,
the evaporation rates $\nu_{\mathcal{A}}$ and $\nu_{\mathcal{B}}$, and the temperature.
No assumptions are made about the internal structure of the diffusiophoretic particles, which could be of various types \cite{Menzel2015}:
The particles could be biological microbes sensing chemoattractants or repellents \cite{Berg1973,Berg1975,Kearns2010}, or synthetic particles like,
for instance, colloidal Janus-particles \cite{Walther2008,Erb2009}.
However, the particles do not need to have a complicated internal structure in order to satisfy the minimal requirements of our model.
Action-reaction symmetry breaking of the diffusiophoretic forces can emerge from various possible asymmetries of transport
coefficients related to the diffusiophoretic particles of type $A$ and $B$ or the chemical species of type $\mathcal{A}$ and $\mathcal{B}$:
As easily seen from Eqs.~\eqref{eq:distinctconcfieldA}, \eqref{eq:diffu_force_on_A} and \eqref{eq:diffu_force_on_B},
it is sufficient to have non-reciprocal diffusiophoretic mobilities, such that $\mu_{A\mathcal{B}} \neq \mu_{B\mathcal{A}}$,
or unequal source terms $s_A \neq s_B$, diffusion coefficients $D_{\mathcal{A}} \neq D_{\mathcal{B}}$, or evaporation rates $\nu_{\mathcal{A}} \neq \nu_{\mathcal{B}}$.

\section{Many-body theory for the pair correlation functions}
\label{sec:paircorrelation}
On the coarse-grained time scale at which the Langevin equation is valid,
an overdamped complex liquid is fully described by the many-body distribution function \PP.
Often times one is interested in more accessible quantities like the pair distribution functions $g^{\ga\gb}(\br, \br^{\prime})$,
which, for the equimolar suspensions studied here, can be defined in the limit $N \to \infty$ in terms of the following $(2N-2)$-fold integrals over $\PP$:
\begin{eqnarray*}
\frac{\rho^2 g^{AA}(\br, \br^\prime)}{N(N-1)} &=& \left\langle \shortint\D \br_3^A \scriptstyle\cdots\displaystyle\shortint\D \br_N^A\shortint\D \br_1^B\scriptstyle\cdots\displaystyle\shortint\D \br_N^B \Psi(\bR, t) \right\rangle, \\
\frac{\rho^2 g^{AB}(\br, \br^\prime)}{N^2}    &=& \left\langle \shortint\D \br_2^A\scriptstyle\cdots\displaystyle\shortint\D \br_N^A\shortint\D \br_2^B\scriptstyle\cdots\displaystyle\shortint\D \br_N^B \Psi(\bR, t) \right\rangle, \\
\frac{\rho^2 g^{BB}(\br, \br^\prime)}{N(N-1)} &=& \left\langle \shortint\D \br_1^A\scriptstyle\cdots\displaystyle\shortint\D \br_N^A\shortint\D \br_3^B\scriptstyle\cdots\displaystyle\shortint\D \br_N^B \Psi(\bR, t) \right\rangle, \\
\end{eqnarray*}
with $\rho = 2N / V$ where $V$ is the suspension volume in case of three-dimensional (3D) systems, or the suspension area in case of two-dimensional (2D) systems.
Alternatively, $ g^{\ga\gb} (\br)$ can be written as
\begin{equation}
 g^{\ga\gb} (\br) =
\frac{V}{N^2}
\left\langle \sum_{i=1}^N\hspace{-.5em}
\sum_{\substack{j=1\\j \neq i \vee \ga \neq \gb}}^N
\delta \left(\br - \br^\ga_i (t)+ \br^\gb_j (t)  \right) \right\rangle.
\label{gofr}
\end{equation}
For an isotropic and homogeneous system $g^{\ga\gb}(r)$ is a function of particle distance only.
The triplet distribution function $g_3^{\ga\gb\gc}(\br, \br^{\prime}, \br^{\prime\prime})$, which is a $(2N-3)$-fold integral over \PP, is  analogously defined~\cite{Hansen2006}.

We start our analysis of particle correlations with the Smoluchowski equation
\begin{align}
 \ddt{\Psi} = \sum\limits_{\alpha = A,B} \frac{1}{\xi^\ga} \sum\limits_{i=1}^N {\nabla}_i^\ga \cdot \left( \kbt~ {\nabla}_i^\ga \Psi  - \bF_i^\ga~\Psi \right),
 \label{smoluchowski}
\end{align}
which is stochastically equivalent to Eq.~\eqref{langevin}, and where ${\nabla}_i^\ga$ is the Nabla operator that differentiates with respect
to the particle position $\boldsymbol{r}_i^\alpha$ and the time- and configuration dependence of $\Psi$ has been dropped for clarity.
Using a $(2N-2)$-fold integration, we transform Eq. \eqref{smoluchowski} into an equation for the pair distribution function~\cite{Kohl2012}.
This equation, however, does not only depend on the pair correlations, but also on the triplet correlations which, in turn, depend on the quadruplet correlations and so on.
We truncate this Bogoliubov–Born–Green–Kirkwood–Yvon (BBGKY) hierarchy~\cite{Lowen2012} using the Kirkwood-superposition approximation~\cite{Kirkwood1935}
\begin{equation}
 g_3^{\ga\gb\gc} (\br, \br^{\prime}, \br^{\prime\prime}) 
 \approx  g^{\ga\gb} (\br, \br^{\prime})~  g^{\ga\gc} (\br,  \br^{\prime\prime}) ~ g^{\gb\gc} (\br^{\prime}, \br^{\prime\prime}). 
\label{kirkwood}
\end{equation}
In our derivation we use the fact that $g^{AB}(r) = g^{BA}(r)$, which is apparent from Eq.~\eqref{gofr}.
The final set of coupled integro-differential equations for $g^{AA}(r)$, $g^{AB}(r)$ and $g^{BB}(r)$ reads
%
\begin{eqnarray}
\left(\frac{\kbt}{\xi^\ga}+\frac{\kbt}{\xi^\gb}\right) \nabla^2 ~ g^{\ga\gb} =
- \hspace{-1em} \sum\limits_{\substack{(\gas, \gbs) = \\ (\ga,\gb),~(\gb,\ga)}} \hspace{-1em}
~\frac{1}{\xi^{\gas}}\nabla \cdot \left[ \rule[-5mm]{0mm}{1.2cm} \bF^{\gas\gbs}~g^{\ga\gb}\right. \nonumber\\
+ \left.\frac{\rho}{2} ~ g^{\ga\gb} \sum_{\gc = A,B}
 \bF^{\gas\gc} \left(g^{\gas\gc} *  g^{\gbs\gc}\right) \rule[-5mm]{0mm}{1.2cm}\right],\hspace{2em}\label{finaleq} 
\end{eqnarray}
where $g^{\ga\gb} \equiv g^{\ga\gb} (r)$ and $F^{\ga\gb}\equiv F^{\ga\gb}(r)$.
In Eq.~\eqref{finaleq},  $(f * g)(r)$ denotes the $d$-dimensional convolution of two isotropic
functions $f(r)$ and $g(r)$, defined as 
\begin{equation}
(f * g)(r) \equiv \int d^d \boldsymbol{r}' f(r^\prime) g (|\br - \br^\prime|). \nonumber
\end{equation}
Eq.~\eqref{finaleq} cannot be solved analytically for nonzero density or non-vanishing force.
We therefore solve it numerically, using fixpoint iteration algorithms \cite{Heinen2014}.
With a double integration, we eliminate the Laplace operator and the divergence.
We solve the convolutions in Fourier space, making use of the FFTLog algorithm for the $d$-dimensional Hankel transform \cite{Hamilton2000,Talman1978}.

To test the accuracy of the approximate theory, we perform Brownian dynamics simulations for two and three dimensions.
We use the forces derived in Eqs.~\eqref{eq:diffu_force_on_A} and \eqref{eq:diffu_force_on_B}, for both the 2D and the 3D case.
The former corresponds to particles that are confined to move in a 2D plane, while the surrounding solvent is
fully three-dimensional. In our model system, the chemical substances $\mathcal{A}$ and $\mathcal{B}$ are free to diffuse throughout the 3D
solvent, irrespective of whether or not the diffusiophoretic, mesoscopic particles are confined to a 2D plane.
For the sake of symmetry and in order to better isolate the effects of nonreciprocity, we consider a system where $\bF^{AA} = \bF^{BB}$ and $\xi^A = \xi^B \equiv \xi$.
We express our simulation parameters in terms of the thermal energy $\kbt$, the number density $\rho$ and the friction coefficient $\xi$, choosing
the Brownian time scale $\tau_B =  \rho^{-2/d}\xi / \kbt$ as a unit of time.

We limit our study to the case of a $r$-independent parameter $\Delta$, which is the case in our model system if $ D_{\mathcal{A}} / \nu_{\mathcal{A}} =   D_{\mathcal{B}} / \nu_{\mathcal{B}}$.
As in our discussion of the Green's function method in Sec.~\ref{sec:model}, we have $\lambda = \sqrt{ D_{\mathcal{A}} / \nu_{\mathcal{A}} }$, which is now a unique
exponential screening length for the concentration profiles of both chemical species $\mathcal{A}$ and $\mathcal{B}$. 
We quantify the strength of the interactions by a constant $\Gamma$, satisfying the equations
\begin{align*}
\frac{\mu_{A\mathcal{A}} ~s_A}{ 4\pi \kbt D_{\mathcal{A}} }   &=  \frac{ \Gamma }{\rho^{1/d} } & 
\frac{\mu_{A\mathcal{B}} ~s_B}{ 4\pi \kbt D_{\mathcal{B}} }   &=  \frac{(1+\Delta)\hspace{0.25em} \Gamma}{\rho^{1/d}}, \\
\frac{\mu_{B\mathcal{A}} ~s_A}{ 4\pi \kbt D_{\mathcal{A}} }   &=  \frac{(1-\Delta)\hspace{0.25em} \Gamma}{\rho^{1/d}} &\hspace{-0.5em}\text{and}~~ 
\frac{\mu_{B\mathcal{B}} ~s_B}{ 4\pi \kbt D_{\mathcal{B}} }   &=  \frac{ \Gamma }{\rho^{1/d} }.
\end{align*}
A potential energy in the usual sense cannot be defined in the general case $\Delta \neq 0$.
However, for the special case of reciprocal interactions ($\Delta=0$), the interactions
above can be described by a pairwise additive potential energy $V(r) = \Gamma \kbt \rho^{-1/d}\exp(-r / \lambda) / r $.
Using a simple forward time-step algorithm~\cite{Colberg2011}, with a time-step $\delta t$ and $ 2 \cdot 10^6$ iterations,
we solve the Langevin Eq.~\eqref{langevin} for $2 \times 20,000$ particles in a 2D quadratic or 3D cubic simulation box with periodic boundary conditions.
The forces are set equal to zero, when the distance between two particles exceeds $5 /\rho^{1/d} $.
The particles are initialized at random positions throughout the system, and the setup is given time to relax.
We monitor the average forces $\langle |\bF ^\ga _i (\bR,t)| \rangle$ and measure that for all of our simulations this value first reduces, and eventually reaches a time-independent steady state.
Then, we calculate the pair and triplet correlations by averaging over multiple snapshots at different times.
\begin{figure}
\centering
\includegraphics[width=\columnwidth]{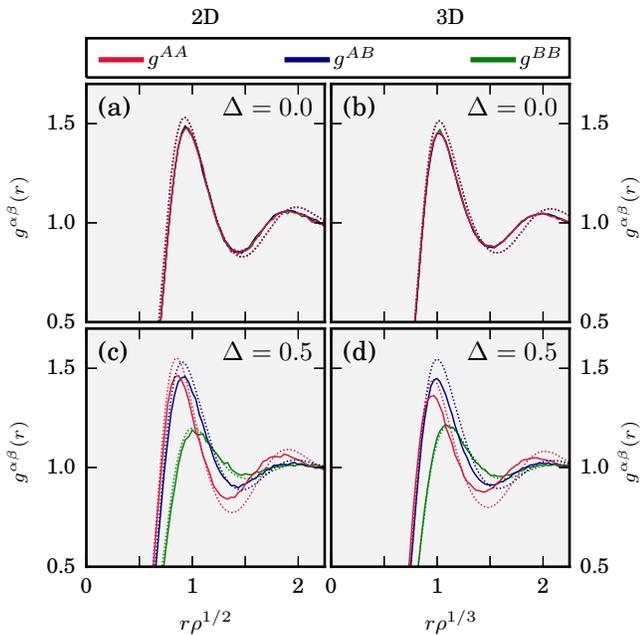}
\caption{(color online) Pair distribution functions $g^{\ga\gb}(r)$ from computer simulation (solid) and theory (dotted).
Panels on the left (a, c) are for $d=2$ spatial dimensions, and panels on the right (b, d) are for $d=3$.
The upper panels (a,b) correspond to the reciprocal case $\Delta = 0.0$.
A nonreciprocal case with $\Delta = 0.5$ is shown in the bottom panels (c, d). 
The remaining parameters of the simulation are $\lambda \rho^{1/d}  = 1/4$ and  
(a) $\Gamma = 100 / 3$, $\delta t/ \tau_B = 3\cdot 10^{-5}$, 
(b) $\Gamma = 200  / 3$, $\delta t/ \tau_B = 1.5\cdot 10^{-5}$,
(c) $\Gamma = 25 $, $\delta t/ \tau_B = 4\cdot 10^{-5}$ and 
(d) $\Gamma = 50 $, $\delta t/ \tau_B = 2\cdot 10^{-5}$.
The plot does not show the region $g(r) < 0.5$, where we observe a very good agreement between theory and simulation.
}
\label{fig1}
\end{figure}
Figure~\ref{fig1} shows a comparison of the pair distribution functions obtained from the theory and the BD simulations for 2D and 3D in the reciprocal and the nonreciprocal case.
For the reciprocal case, $\Delta = 0$, where $g^{AA}(r) = g^{AB}(r) = g^{BB}(r)$, the theory predicts the simulation results for $g^{\alpha\beta}(r)$ with high precision.
In case of nonreciprocal forces, $\Delta = 0.5$, the deviations between theory and simulation results are larger, but the theory maintains a rather good accuracy level and it
continues to capture all qualitative features of the simulation.
Note also, that all functions $g^{\alpha\beta}(r)$ exhibit a pronounced `correlation hole' at small values of $r$, because the repulsive particles almost never come
into close contact. This provides an \textit{a posteriori} justification of our point-particle assumption in Sec.~\ref{sec:model}: Particles that are significantly
smaller in diameter than the correlation hole have a negligible likelihood of direct contact, and can therefore be approximated as point-like. 

Without showing all results here, we have observed both in our simulations and our theory results,
and for 2D as well as for 3D systems,
that the principal peak value, $g^{AA}(r^{AA}_{max})$, of the function $g^{AA}(r)$ can assume a smaller or larger value than the principal
peak $g^{AB}(r^{AB}_{max})$. The peak-height ordering depends on the parameters $(\Gamma, \lambda, \Delta)$
of the nonreciprocal interactions and on the density $\rho$.
\begin{figure}
\centering
\includegraphics[width=\columnwidth]{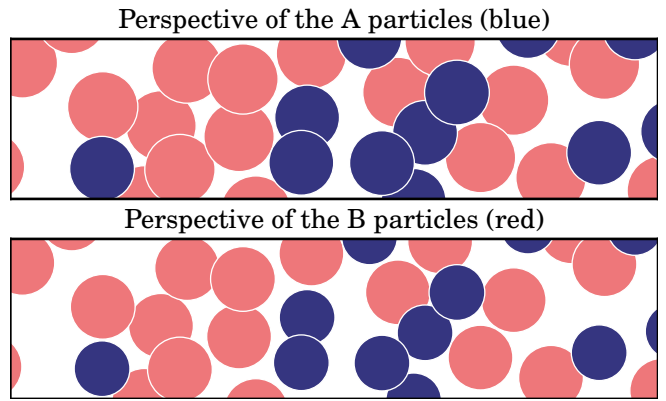}
\caption{(color online) Typical snapshot for a $2D$-simulation with $\Delta= 0.5$ and  $\Gamma = 25$.
The system exhibits different effective densities for $A$ and $B$ particles.
The radii of the plotted disks are proportional to the effective radii $r_{\rm eff}^{\ga\gb}$.
}
\label{fig:snapshot}
\end{figure}
In our simulation and theory results we observe that the principal peak height of function $g^{BB}(r)$ is always less than the peak heights of both functions $g^{AB}(r)$ and $g^{AA}(r)$.
For an intuitive understanding of the less pronounced peak in $g^{BB}(r)$, let us introduce effective radii $r_{\rm eff}^{\ga\gb}$
via the condition that $|\mathbf{F}^{\ga\gb}(r_{\rm eff}^{\ga\gb}) |=  \kbt / \lambda$.
In \fig{fig:snapshot} we show a snapshot from a 2D system with nonreciprocal interactions twice,
using different effective radii for the plotted disks that are centered around the particle positions $r_i^A$ (blue) and $r_i^B$ (red): 
In the top panel of the figure, the effective radius of the red, $B$-type disks is $r_{\rm eff}^{BA}$, and 
in the bottom panel, the effective radius of the blue, $A$-type disks is $r_{\rm eff}^{AB}$, which is less than $r_{\rm eff}^{BA}$.
The same effective radius, $r_{\rm eff}^{AA} = r_{\rm eff}^{BB}$, is used for the blue disks in the upper panel and for the red disks in the lower panel.
Clearly, the system is effectively more crowded for the $A$-type particles than for the $B$-type particles, which explains the weaker principal peak in $g^{BB}(r)$.

\section{Kirkwood approximation for nonreciprocal interactions}

The approximation that allows us to solve the many-body Smoluchowski equation numerically is the Kirkwood superposition in Eq.~\eqref{kirkwood}.
In case of thermodynamic equilibrium, it is known how this approximation breaks down at high density \cite{Barrat1987,Barrat1988,Lowen1992}.
In the following, we test the Kirkwood superposition approximation in case of the nonequilibrium steady state
of Brownian suspensions with nonreciprocal interactions, by comparison to our highly accurate computer simulation data.

\begin{figure}
\centering
\includegraphics[width=.65\columnwidth]{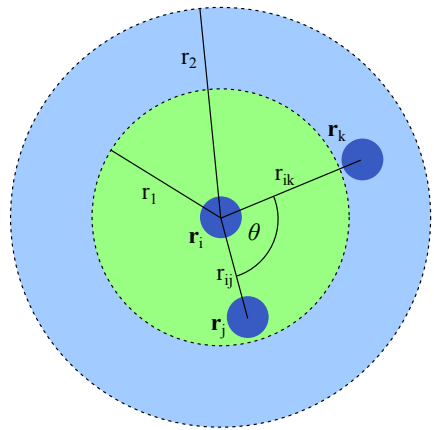}
\caption{(color online) The bond angle distribution function  $g_3 ( \theta, r_1, r_2)$ characterizes triplets of particles with the inter-particle distances $r_{ij} < r_1$ and $r_{ik} < r_2$ by the bond angle theta $\theta$.
 }
\label{sketchbondangle}
\end{figure}

One way to visualize a projection of the triplet correlation function $g_3^{\ga\gb\gc}(\br, \br^{\prime}, \br^{\prime\prime})$ is via the
bond angle distribution function $g_3 ( \theta, r_1, r_2)$~\cite{Lowen1992}.
This function characterizes triplets that have one inter-particle distance smaller than $r_1$ and another inter-particle distance smaller than $r_2$,
by the bond angle $\theta$ between the two straight lines that connect the particle centers (see \fig{sketchbondangle}).
The function $g_3 ( \theta, r_1, r_2)$ is normalized such that the integral over all bond angles yields unity.
Often, $r_1$ and $r_2$ are chosen as the first minimum of the pair distribution function.
However, this is not uniquely defined for a binary mixture.
To avoid this ambiguity, we choose the parameters as the first minimum of a corresponding $g^{AA}(r)$ for a simulation with $\Delta = 0$, which we call $R$,
and which should not be confused with the norm of the super vector $\bR$.
For 2D and strong particle interactions, pronounced peaks around values of $\theta$ that are integer multiples of $60^{\circ}$ indicate triangular short-range order of the liquid \cite{Lowen1992}.

Assuming Kirkwood superposition we can approximate the bond angle distribution function trough a combination of pair distribution functions.
We define the unnormalized Kirkwood-approximation $G_{3,K}^{\ga\gb\gc}(\theta, r_1, r_2)$ for the bond angle distribution function in 2D as
\begin{equation}
\begin{split}
G_{3,K}^{\ga\gb\gc} ( \theta, r_1, r_2) \equiv \int_0^{r_1}\hspace*{-.7em}\int_0^{r_2}\hspace*{-.7em} \D r& \D r^\prime ~r ~ r^\prime g^{\ga\gb}(r) g^{\ga\gc}(r^\prime)
\\&\times   g^{\gb\gc}(\sqrt{r^2 + {r^\prime}^2- 2 r r^\prime \cos\theta})
\end{split} \nonumber
\end{equation}  
and similarly in 3D as
\begin{equation}
\begin{split}
G_{3,K}^{\ga\gb\gc} ( \theta, r_1, r_2) \equiv \sin \theta \int_0^{r_1}&\hspace*{-.7em}\int_0^{r_2}\hspace*{-.7em} \D r \D r^\prime ~r^2 ~ {r^\prime}^2 g^{\ga\gb}(r) g^{\ga\gc}(r^\prime)
\\&\times   g^{\gb\gc}(\sqrt{r^2 + {r^\prime}^2- 2 r r^\prime \cos\theta}).
\end{split} \nonumber
\end{equation}
Applying normalization we arrive at the Kirkwood approximation of the bond angle distribution function, 
\begin{equation} 
g_{3,K}^{\ga\gb\gc} (\theta, r_1, r_2) =  \frac{\displaystyle{G^{\ga\gb\gc}_{3,K} ( \theta, r_1, r_2)}}{\displaystyle{\int_0^\pi \D \theta ~ G^{\ga\gb\gc}_{3,K} ( \theta, r_1, r_2)}}. 
\label{kirkwood_bondangle}
\end{equation}
\begin{figure*}
\includegraphics[scale=1]{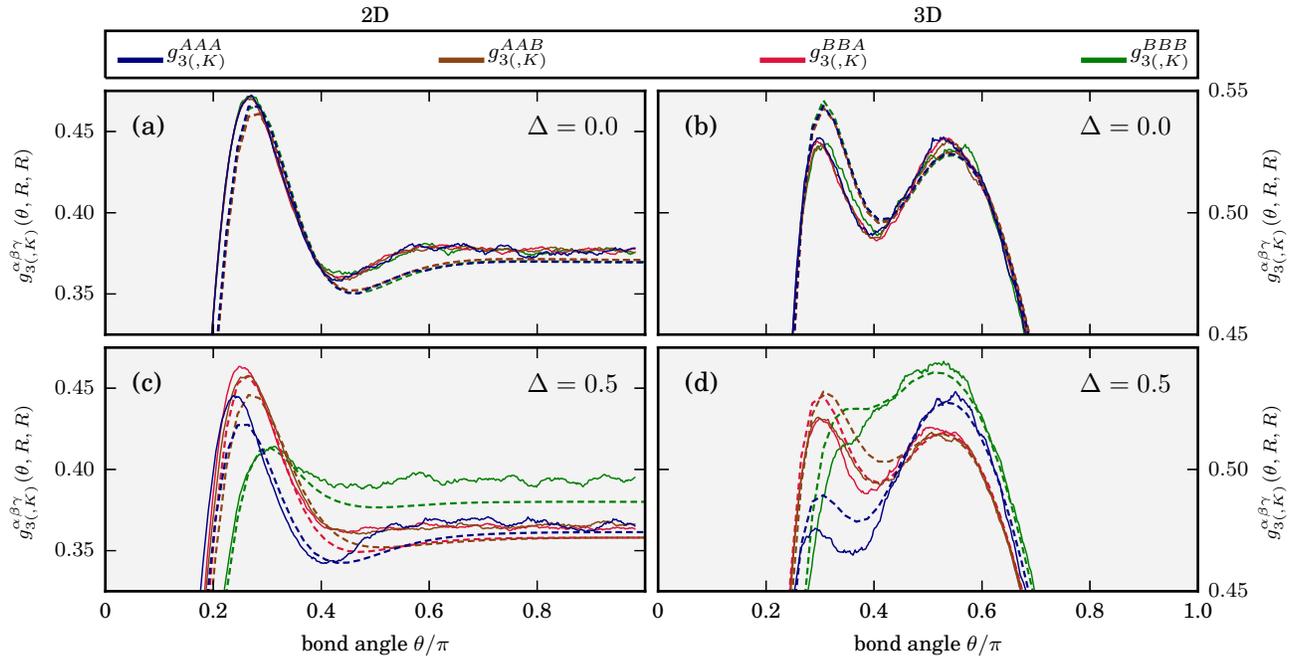}
\caption{(color online) Bond angle distribution function $g^{\ga\gb\gc}_3(\theta, R,  R )$ computed directly from our computer simulation (solid curves)
and Kirkwood approximations, $g^{\ga\gb\gc}_{3,K}(\theta, R,  R )$, of the bond angle distribution functions (dashed curves).
The functions $g^{\ga\gb\gc}_{3,K}(\theta, R,  R )$ are computed on basis of Eq.~\eqref{kirkwood_bondangle}, using the pair correlation
functions $g^{\alpha\beta}(r)$ from the simulations as input. Simulation parameters are the same as for Fig.~\ref{fig1}.
}
\label{fig_triplets}
\end{figure*}
In Fig.~\ref{fig_triplets} we plot the functions $g^{\ga\gb\gc}_3(\theta, R,  R )$ and $g^{\ga\gb\gc}_{3,K}(\theta, R,  R )$, both extracted from our simulations.
As in Fig.~\ref{fig1}, we show data for the 2D and 3D case, both for reciprocal and nonreciprocal interactions. 
All simulated systems are clearly in the liquid state, as signaled by the very gentle principal peak at a bond angle $\theta$ just below $\pi / 3$.
Low values of the bond angle distribution functions at small values of $\theta$ correspond once again to a correlation hole:
It is very unlikely for a pair of repulsive particles to occupy the same space. 
In the 3D case, large bond angles are also untypical.
The probability of finding a particle at a given angle scales with the solid angle in $3D$, which is proportional to $\sin \theta$. 
For 2D systems, the bond angle distribution functions at angles larger than $\pi /2$ are almost constant.

We find that the Kirkwood approximation is very accurate for the studied systems with reciprocal interactions,
and somewhat less accurate in case of systems with nonreciprocal interactions.
As expected, the discrepancies between $g^{\ga\gb\gc}_{3,K}(\theta, R,  R )$ and $g^{\ga\gb\gc}_{3}(\theta, R,  R )$ are strongest for those systems
where the pair distribution functions from the many-body Smoluchowski theory with Kirkwood closure exhibit the lowest level of accuracy (\textit{c.f.}, Figs.~\ref{fig1},\ref{fig_triplets}).

\section{Conclusions}

We have studied 2D and 3D systems of Brownian particles with reciprocal and nonreciprocal particle interactions.
A microscopic theory based on the many-body Smoluchowski equation 
with the Kirkwood superposition approximation as a closure predicts the particle pair-correlation functions with good accuracy.
Nonreciprocal interactions have distinct influence of the pair-correlations, as revealed by the differences between the correlation functions for systems with reciprocal and nonreciprocal forces.
Our predictions for the pair- and triplet-correlation functions can be tested experimentally with binary mixtures of diffusiophoretic particles.

Future theory could improve the closure beyond the Kirkwood superposition principle. 
Possible candidates for future development are dynamical density functional theory 
\cite{Marconi1999,Archer2004,Rex2007,Espanol2009,Wensink2008} or
mode coupling theory \cite{Szamel2015} for nonequilibrium systems,
which still need to be generalized to systems with nonreciprocal interactions.
Furthermore, the effect of different non-reciprocity classes (constant versus $r$-dependent $\Delta$) on the structural correlations should be carefully explored.

\begin{acknowledgments}
This work was supported by the ERC Advanced Grant INTERCOCOS (Grant No. 267499)  and from the Russian Scientific Foundation, Project No.~14-43-00053.
M.~H. acknowledges support by a fellowship within the Postdoc-Program of the German Academic Exchange Service (DAAD).
The authors thank Matthias Kohl for helpful discussions.
\end{acknowledgments}

\bibliographystyle{iopart-num}
\providecommand{\newblock}{}

\end{document}